\documentclass[a4paper,11pt]{article}
\usepackage{jinstpub} 
\usepackage{lineno}

\newcommand{\jpsi}{J/\psi}

\title{\boldmath Search for Charged Lepton Flavor Violation in $\jpsi$ decays at BESIII}

\author{Jingshu Li, Zhengyun You}
\affiliation{School of Physics, Sun Yat-Sen University,\\
Guangzhou 510275, China}

\emailAdd{lijsh53@mail2.sysu.edu.cn}
\emailAdd{youzhy5@mail.sysu.edu.cn}

\abstract{The observation of any charged lepton flavor violation (CLFV) process would be a clear signal of new physics beyond the Standard Model. Various decay modes, including lepton ($\mu, \tau$) decays, pseudoscalar meson ($K,\pi$) decays, vector meson ($\phi, \jpsi, \Upsilon$) decays, and Higgs decays, have been explored to detect CLFV. Focusing on the search for CLFV at the $\tau$-charm region, the results of the search for $\jpsi\to e\tau / e\mu$ using the 10 billion $\jpsi$ events collected by the BESIII experiment are presented. The upper limits (ULs) at the 90$\%$ confidence level are  $\mathcal{B}(\jpsi\to e^\pm\tau^\mp)<7.5\times10^{-8}$ and  $\mathcal{B}(\jpsi\to e^\pm\mu^\mp)<4.5\times10^{-9}$, respectively. Improving the previously published limits by two orders of magnitudes, the results are the most stringent CLFV searches in heavy quarkonium system.}


\arxivnumber{1234.56789} 

\begin{document}
\maketitle
\flushbottom

\section{Introduction}
\label{sec:intro}

The lepton flavor violating (LFV) decays are forbidden in the Standard Model (SM), however, neutrinos with extremely small masses can transform their flavors into another via oscillation effect, which reveals that the neutral lepton flavor is not conserved. For charged leptons, no LFV decay has been observed so far. Therefore, the observation of any LFV decay would be a clear signal of new physics (NP) beyond the SM~\cite{Bern:2013, Cei:2014}.

\begin{figure}[htbp]
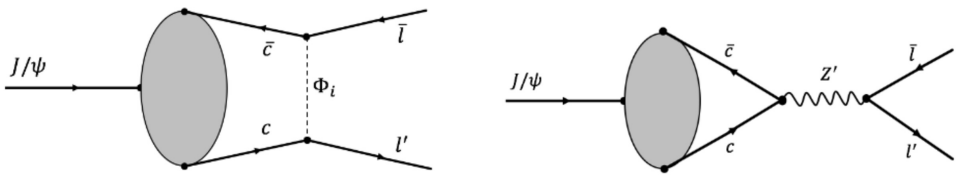

\centering
\includegraphics[width=0.38\textwidth]{figure/leptoquark.pdf}
\qquad
\includegraphics[width=0.396\textwidth]{figure/topcolor.pdf}
\caption{Diagrams of LFV decay $\jpsi\to l^{'}\bar{l}$ via leptoquark (left) and $Z^{'}$ (right).\label{fig:fey}}
\end{figure}

Various theoretical models predict CLFV, including the two Higgs doublet model with extra Yukawa couplings~\cite{Hou:2021zqq}, a model that constitutes a simple example of tree-level off-diagonal Majorana couplings not suppressed by neutrino masses~\cite{Escribano:2021uhf}, Supersymmetry (SUSY) with vector-like leptons and a right-handed neutrino~\cite{Kitano:2000, Borzu:1986},  the minimal SUSY model with gauged baryon number and lepton number~\cite{msy}, etc. Figure~\ref{fig:fey} shows the diagram of LFV decay $\jpsi\to l^{'}\bar{l}$ via leptoquarks or $Z^{'}$ in the topcolor assisted technicolor (TC2) models~\cite{yue:2000zp}.

Experimentally, CLFV decays of leptons ($\mu, \tau$), pseudoscalar mesons ($K, \pi, B$), and vector mesons ($\phi, \jpsi, \Upsilon$) are studied with massive data sets. The BES collaboration obtained ULs of $\mathcal{B}(\jpsi\to e\mu)<1.1\times10^{-6}$~\cite{bes:2003}, $\mathcal{B}(\jpsi\to e\tau)<8.3\times10^{-6}$, and $\mathcal{B}(\jpsi\to \mu\tau)<2.0\times10^{-6}$~\cite{bes:2004} by analyzing $58\times10^{6}~\jpsi$ events. In this paper, searches for the CLFV process $\jpsi\to e^{\pm}\tau^{\mp}$ with $\tau^{\mp}\to \pi^{\mp}\pi^{0}\nu_{\tau}$ using $10\times 10^{9}$ $\jpsi$ events and $\jpsi\to e^{\pm}\mu^{\mp}$ using $8.998\times 10^{9}$ $\jpsi$ events collected with the BESIII detector are reported.

\section{BESIII detector}

The BESIII detector~\cite{Ablikim:2009aa, detvis} records symmetric $e^+e^-$ collisions  provided by the BEPCII storage ring~\cite{Yu:IPAC2016-TUYA01}, which operates with a peak luminosity of $1\times10^{33}$~cm$^{-2}$s$^{-1}$
in the center-of-mass energy range from 2.0 to 4.95~GeV.
BESIII has collected large data samples in this energy region~\cite{Ablikim:2019hff}. 

\section{CLFV results from BESIII}

The analyses are based on $\jpsi$ events collected in the years of 2009, 2012, 2018, and 2019 at $\sqrt{s}=3.097~\rm{GeV}$ at BESIII ($\jpsi\to e^{\pm}\mu^{\mp}$ without 2012)~\cite{bes3:totJpsiNumber2}. 
Inclusive MC events are used to study the backgrounds from $\jpsi$ decays and the signal process are simulated with PHOTOS and the VLL decay model in EVTGEN~\cite{ref:evtgen1}. A semi-blind analysis method with $10\%$ of the full $J/\psi$ data randomly selected is used to avoid involuntary bias during the analyses~\cite{bes:2021, bes:2022}.

\subsection{Search for CLFV decay $\jpsi\to e\tau$}

\begin{figure}[htbp]
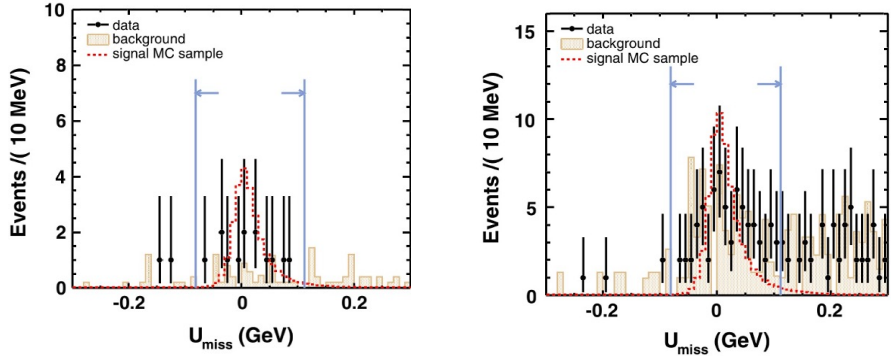

\centering
\includegraphics[width=.35\textwidth]{figure/umiss1.pdf}
\qquad
\includegraphics[width=.37\textwidth]{figure/umiss2.pdf}
\caption{The $U_{\rm{miss}}$ as well as corresponding background distribution of data sample I (left) and II (right), data sample I for $1310.6\times 10^{6}$ events collected in 2009 and 2012, and data sample II collected in 2018 and 2019.\label{umiss}}
\end{figure}

Since $\tau$ is reconstructed by $\pi^{\mp}\pi^{0}\nu_{\tau}$, we need to select one electron and one charged pion in the final state. Additionally, we need to identify at least two photon showers and one $\pi^{0}$ particle.  The final-state electron resulting from the decay of $\jpsi\to e\tau$ is monochromatic.
The missing energy of the undetected neutrino is calculated by $E_{\rm{miss}}=E_{\rm{CMS}}-E_{e}-E_{\pi}-E_{\pi^{0}}$, where $E_{\rm{CMS}}$ is  the center-of-mass energy of the initial $e^{+}e^{-}$ system and $E_{e}$, $E_{\pi}$ and $E_{\pi^{0}}$ are the energies of the electron, charged pion, and neutral pion in the rest frame of the $e^{+}e^{-}$ system. 
$E_{miss}$ is required to be greater than $0.43 \rm{GeV}$. 
The primary sources of background contamination in this analysis are the continuum process, e.g., radiative Bhabha scattering, and hadronic $\jpsi$ decay modes, e.g. $\jpsi\to \pi^{+}\pi^{-}\pi^{0}$. The $U_{\rm{miss}}$ is defined as $U_{\rm{miss}}=E_{\rm{miss}}-c|P_{\rm{miss}}|$, where $P_{\rm{miss}}=P_{\rm{CMS}}-P_{e}-P_{\pi}-P_{\pi^{0}}$ and $P$ are the corresponding moment. $U_{\rm{miss}}$ is used to extract the signals, the areas between the arrows represent the signal region, as shown in Figure~\ref{umiss}.

Using the Bayesian method~\cite{likelihood}, the UL at the 90$\%$ confidence level (C.L.) for the branching fraction is determined to be $\mathcal{B}(\jpsi\to e^\pm\tau^\mp)<7.5\times10^{-8}$. It is the first submitted paper based on full 10 billion $\jpsi$ data of BESIII, and the result improves the previous published limits by two orders of magnitude and is comparable with the theoretical predictions.

\subsection{Search for CLFV decay $\jpsi\to e\mu$}

This analysis is based on $8.998\times 10^{9}$ $\jpsi$ events. About same size of inclusive MC and 10 times of exclusive MC events are generated to study the background. Each $\jpsi$ candidate is reconstructed with two back-to-back good charged tracks, which will be further identified as electron and muon. 
The signal region is defined based on the conservation of energy and momentum, with the conditions $|\Sigma\vec{p}|/\sqrt{s}\leqslant0.02$ and $0.95\leqslant E_{\rm vis}/\sqrt{s}\leqslant1.04$, which are optimized by maximizing the figure of merit. Here, $|\Sigma\vec{p}|/\sqrt{s}$ represents the magnitude of the vector sum of the momenta normalized to the center of mass energy, and $E_{\rm vis}$ is the total reconstructed energy. 29 candidate events are observed in the signal region upon analyzing the full data, which are consistent with the background estimation using MC data.

\begin{figure*}[htbp]
\centering
\includegraphics[width=0.5\textwidth]{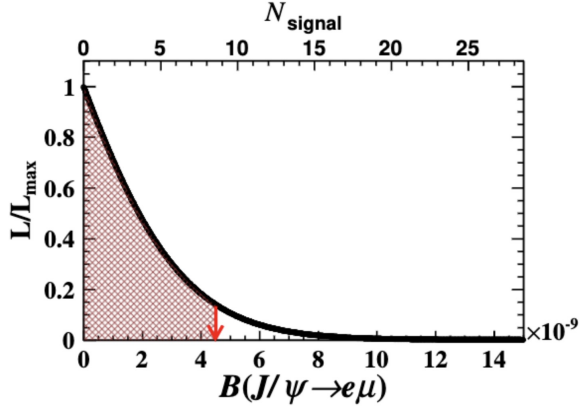}
\caption{Normalized likelihood distribution as a function of the assumed $\mathcal{B}(\jpsi\to e\mu)$. The red arrow points to the position of the UL at the $90\%$ C.L. The upper $x$-axis shows the corresponding values of the signal yields $N_{\rm signal}$. }
\label{fig:likelihood}
\end{figure*}

A likelihood fit is constructed to get the UL of branching fraction, which is set to be $\mathcal{B}(\jpsi\to e^\pm\mu^\mp)<4.5\times10^{-9}$ at the 90$\%$ C.L., as shown in Figure~\ref{fig:likelihood}. This result improves the previously published limits by more than a factor of 30, reaching a level comparable to the theoretical predictions.
\vspace{-0.0cm}
\vspace{-0.0cm}

\section{Summary}

BESIII has collected the largest data samples of $\jpsi$ on threshold in the world. The high statistics $\jpsi$ data provides a unique opportunity for a thorough investigation of CLFV to search for new physics beyond the SM. This paper presents the searches of the CLFV process $\jpsi\to e\tau$ and $\jpsi\to e\mu$ at BESIII, the UL is determined to be $\mathcal{B}(\jpsi\to e^\pm\tau^\mp)<7.5\times10^{-8}$ and $\mathcal{B}(\jpsi\to e^\pm\mu^\mp)<4.5\times10^{-9}$ at the 90$\%$ C.L.. These result can be used to constrain new physics model parameter spaces and the $\jpsi\to e\mu$ result is the most stringent CLFV constraint in the heavy quarkonium sector up to now.




\end{document}